\documentclass[11pt,a4paper]{article}
\usepackage[utf8]{inputenc}
\usepackage[dvips]{graphicx}

\author{Jakub Kowalski, Wojciech Waga, Marta Zawierta, Stanis\l{}aw Cebrat}
\title{Phase transition in the genome evolution favours non-random distribution of genes on chromosomes}

\begin{document}
\maketitle

\begin{abstract}We have used the Monte Carlo based computer models to show that selection pressure could affect the distribution of recombination hotspots along the chromosome. Close to critical crossover rate, where genomes may switch between the Darwinian purifying selection or complementation of haplotypes, the distribution of recombination events and the force of selection exerted on genes affect the structure of chromosomes. The order of expression of genes and their location on chromosome may decide about the extinction or survival of competing populations.
\end{abstract}

\section*{Key words:} Monte Carlo simulation, chromosome structure, evolution, complementation, recombination, competition.

\section{Introduction}

According to neo-Darwinism there are three forces driving the evolution of~biological systems: mutations, recombinations and selection of which the~only directional force is selection favouring the fittest individuals \cite{Ayala}. Nevertheless, there is a very sophisticated interplay between these forces. Selection favours some structures where recombination events as well as mutations are not totally random. They could be highly biased in both, frequency and location. 

\begin{figure}
\includegraphics[angle=0,width=\textwidth]{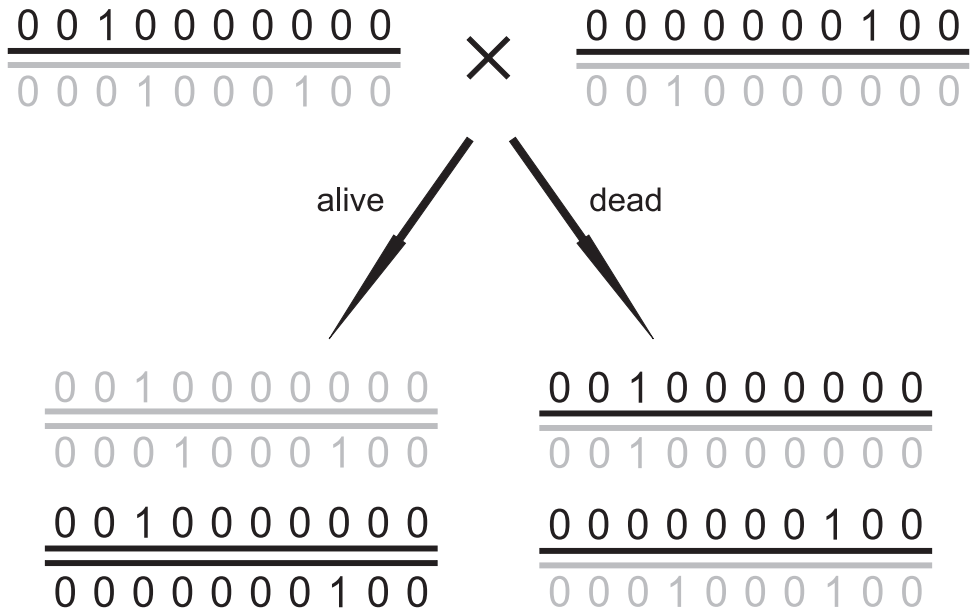}
\caption{The puryfying selection strategy.}\label{f1}
\end{figure}
\begin{figure}
\includegraphics[angle=0,width=\textwidth]{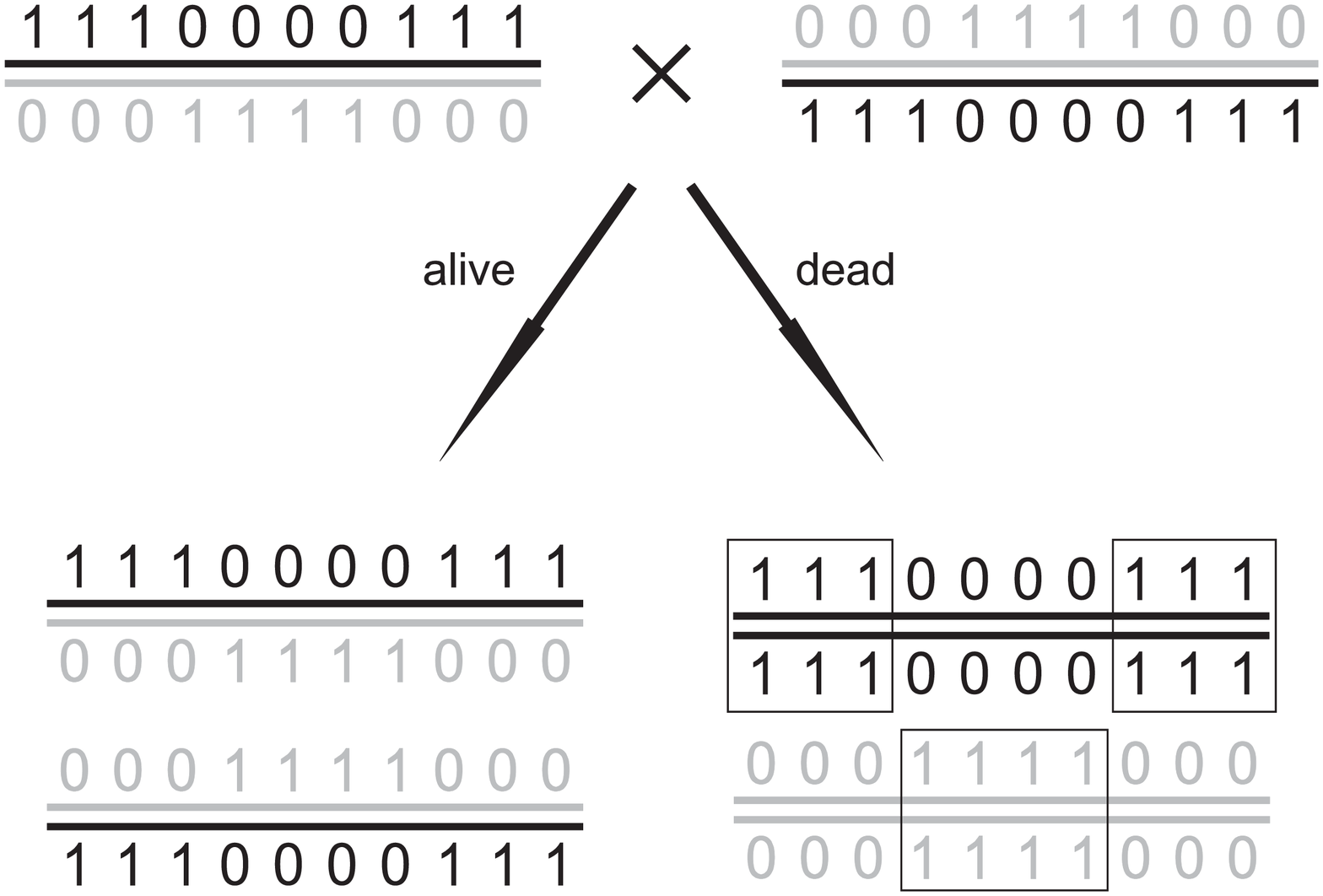}
\caption{The complementing strategy.}\label{f2}
\end{figure}

The Monte Carlo modelling of the phenomena of genome evolution has revealed that depending on the intragenomic recombination rate and effective population size (or more precisely inbreeding coefficient), genomes can switch between two different strategies of evolution --- the Darwinian purifying selection and complementation of haplotypes \cite{Zawierta}, \cite{Waga}. Strategy of~purifying selection is advantageous in the large, randomly mating, panmictic populations with high intragenomic recombination rate, where the genetic relations between sexual partners are very low (so called Mendelian populations). In such situations selection eliminates effectively individuals with homozygous defective loci and the fraction of defective alleles in the genetic pool of population is relatively low. In small populations, the probability of inheriting by an individual the long strings of genes from both parents originated from the common ancestor is higher. Under such condition the~other strategy is more advantageous --- complementing the whole strings of defective alleles by their wild forms. The situation is illustrated in Fig.~\ref{f1} and~\ref{f2}. The complementing strategy enables the accumulation of higher fraction of defective alleles in the genetic pool of population. The second possibility is very often overlooked and even recently some models of biological evolution are built on the base of so called Wright-Fisher (W-F) model \cite{Fisher}, \cite{Wright1}, \cite{Wright2} of Mendelian population, under the assumption of independent inheritance of alleles in different loci \cite{Som}, \cite{Redfield}.

It is obvious that switching between the two strategies depends on the evolutionary costs and reproduction potential of populations. It has been found that such a switching has a character of phase transition \cite{Zawierta2}. For a~given set of evolving populations --- population size, length of chromosomes, mutation rate --- the critical intragenomic recombination rate can be found. Below this recombination rate the complementing strategy is more advantageous while above this critical recombination rate the purifying strategy ensures higher reproduction potential of population. Population size influences the value of critical recombination rate only if the panmixia (random selection of mates) is assumed. In such populations the power law relations between the recombination rate and populations' sizes have been found \cite{Zawierta2}. If some constraints are imposed on the totally random selection of mates in~the~panmictic populations (i.e. distance limit between mates), the value of~critical intragenomic recombination rate depends on the inbreeding coefficient. Modelling the evolution of population on the lattice, when the distances for looking for the mates and for placing the offspring are the constant parameters of the model, the value of critical intragenomic recombination rate does not depend on the population size \cite{Waga}. Furthermore, simulations on lattice revealed that under some conditions restriction imposed on recombination rate increases the population expansion and enables the sympatric speciation \cite{Waga}. 

In this paper we are studying the functional properties of chromosomes in the Penna model under recombination rate close to critical values. The~Penna model has been used in many simulations of age structured populations \cite{Penna}. It is very convenient for such studies because selection values of genes differ in the model depending on the period of life when they are expressed. In particular we have studied the role of distribution of the recombination hot spots along the chromosomes.

\section{Model}

\begin{figure}
\includegraphics[angle=0,width=\textwidth]{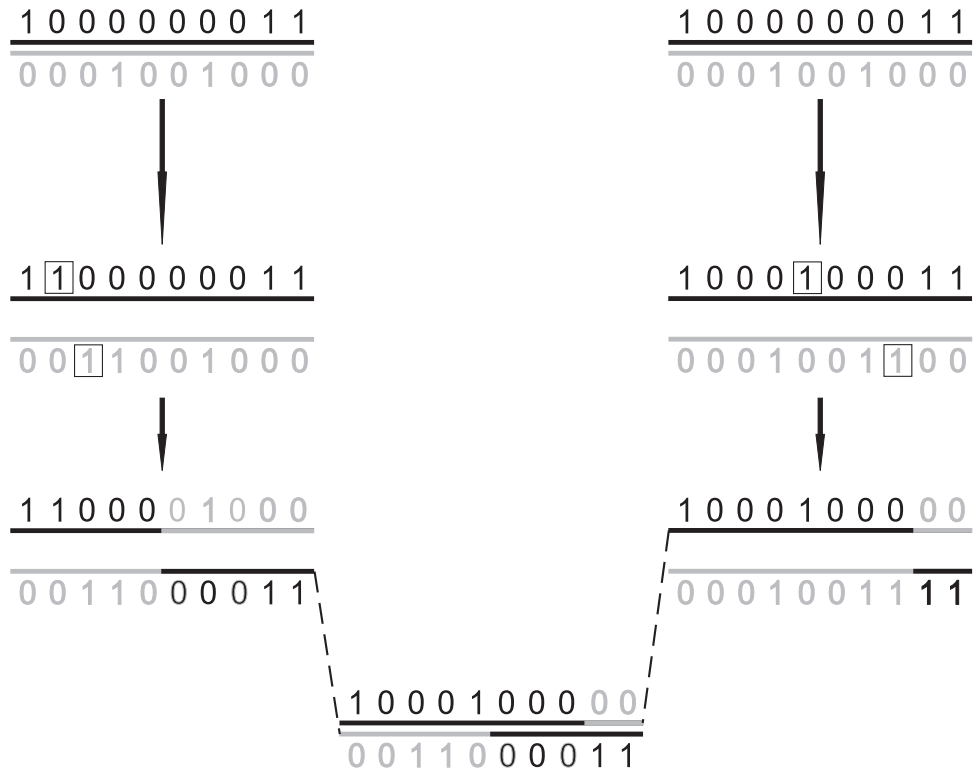}
\caption{The scheme of reproduction process.}\label{f3}
\end{figure}

Detailed descriptions of the standard diploid Penna model can be found in~many original papers \cite{scaling} or reviews \cite{Moss}, \cite{Stauffer}. We have used the modified, diploid version of the Penna model with sexual reproduction. Panmictic populations are composed of N individuals represented by their diploid genomes. Each genome is composed of one pair of haplotypes (bitstrings) L=128 bits long. Bits correspond to genes and, if they are located at the corresponding positions in the bitstrings they correspond to alleles. Bits set to 0 correspond to the wild - correct alleles, bits set to 1 correspond to the defective alleles. Defective phenotype is determined only if both bits at the corresponding positions in the bitstrings are set to 1. Bits (genes) are switched on (expressed) chronologically. In the standard Penna model, during each Monte Carlo (MC) step one bit in each bitstring is switched on; in the first step the first bit, in the second step the second bit and so on. In our version the sequence of switching on the bits can be modified. It will be described in details in the next section. Anyway, the number of switched on bits determines the age of individual. When the individual reaches the age R=80 it can reproduce. In one MC step, each female at the reproduction age can give birth to B offsprings. She copies her genome introducing a new mutation into each haplotype in the random position with probability M. If the~bit chosen for mutation is 0 it is changed for 1, if it is 1 it stays 1 (there are no reversions). Two copied and mutated bitstrings are paired and recombined in the random position with probability C in the process mimicking crossover. One product of recombination of each pair of chromosomes is randomly assorted to the female gamete. Then, the female looks for a male partner at the reproduction age, who produces male gamete on the same way. Male and female gametes fuse forming a diploid genome of a newborn. Its first bits will be switched on in the next MC step. The male individual, after reproduction with one female, returns to the pool of males at the reproduction age and it can produce an offspring with other female(s) in~the~same MC~step. The~scheme of the process is shown in Fig. \ref{f3}. As mentioned above, if both bits in corresponding position are set to 1 (are defective) they determine the deleterious phenotype when switched on. If the~number of~switched~on defective phenotypes reaches the declared threshold T --- the~individual dies because of genetic death. The size of panmictic populations is controlled by Verhulst factor operating in our version of model only at~birth, as proposed by \cite{SaMartins}. The Verhulst factor $V=1-N_{t}/N_{max}$, where; $V$ -- probability of the newborn survival, $N_{t}$ -- the actual size of population and Nmax is called sometimes the maximum capacity of the environment. 
\section{Genome evolution far from the critical point of recombination}

\begin{figure}
\includegraphics[angle=0,width=\textwidth]{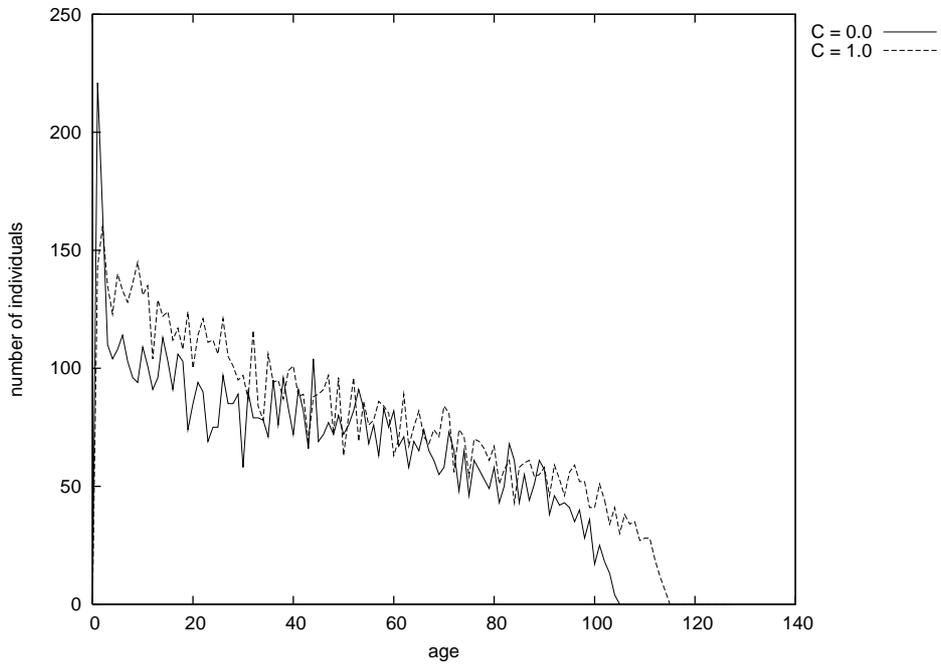}
\caption{The age structure of populations.}\label{f4}
\end{figure}
\begin{figure}
\includegraphics[angle=0,width=\textwidth]{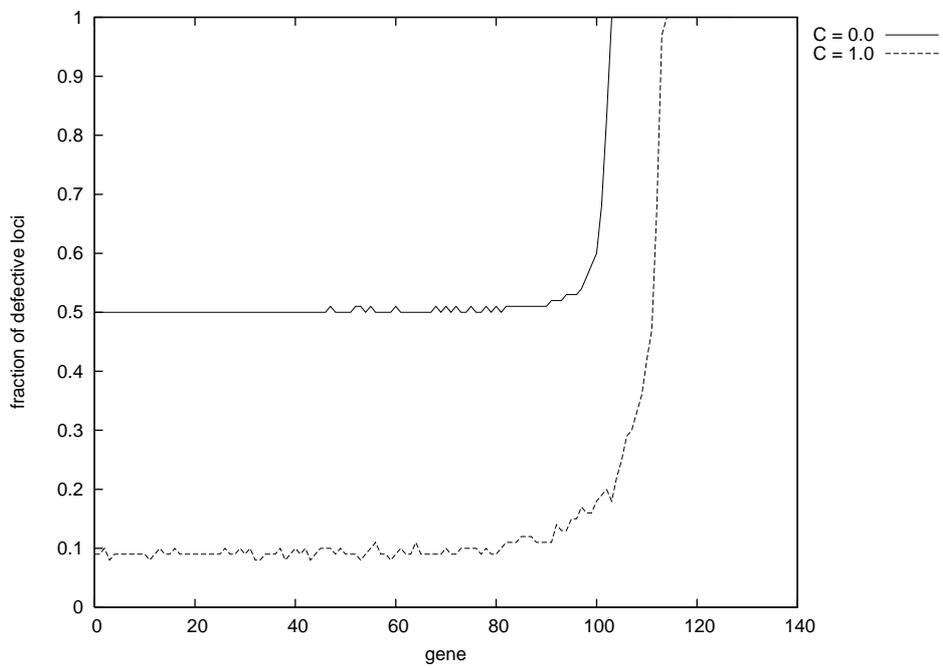}
\caption{The genetic structure of populations.}\label{f5}
\end{figure}

In Fig. \ref{f4} and \ref{f5} the age and the genetic structure of populations simulated with the standard diploid Penna model are shown. Parameters of simulations were: one pair of bitstrings, L=128, $N_{max}$=10~000, R=80, M=1, T=1, B=2 and C=1 or C=0. The results show that populations evolving under recombination rate 1 are larger, they have lower fraction of defective alleles in~the~part of~genomes expressed before the reproduction age (up to 80), larger fraction of individuals in the reproduction age and higher maximum lifespan. Nevertheless, populations evolving without crossover survived, though they had very high fractions (0.5) of defective genes expressed before the minimum reproduction age. If we assume that the defective genes are distributed randomly in the pool of haplotypes, the average probability of surviving the~individual until the~minimum reproduction age is of the order of 4~x~$10^{-8}$. Under such conditions, populations should be extinct. In fact, in the whole genetic pool only two different and complementing haplotypes exist. The effect of accumulation of defects in the diploid version of the Penna model was observed previously \cite{Aga}. Furthermore, it has been also found that gamete recognition --- the process in which complementing gametes preferentially form zygotes --- increases the reproduction potential of populations \cite{Cebrat}. 

Intuitively, it seems that under fixed parameters of simulations the only condition which influences the fraction of~defective alleles is the~period of~life when the genes are expressed, it should not depend on the order of genes along the chromosome and its genetic map.  To check this we have inverted the second part of chromosomes (bits from 42 to 128). Now, bit expressed in~the~$42^{nd}$ MC step is located at the end of chromosome. We have simulated the evolution of populations with the standard order of bits (not inverted) and with the inverted second part of bitstrings under the same conditions (the same values of the rest of parameters). The parameters describing the obtained populations (size of populations, fraction of defective genes, maximum lifspan) suggest that the structure of chromosomes --- the co-linearity or the lack of co-linearity between the sequence of expression and physical sequence of genes on chromosomes do not influence the evolutionary value of population at least under the studied crossover rate: C=1 and C=0. Nevertheless, in the next sections it will be shown that the fitness of the two populations with different structures of chromosomes are equivalent only far from critical values of intragenomic recombination rate.

\section{The role of chromosome structure close to the critical point of recombination}

\begin{figure}
\includegraphics[angle=0,width=\textwidth]{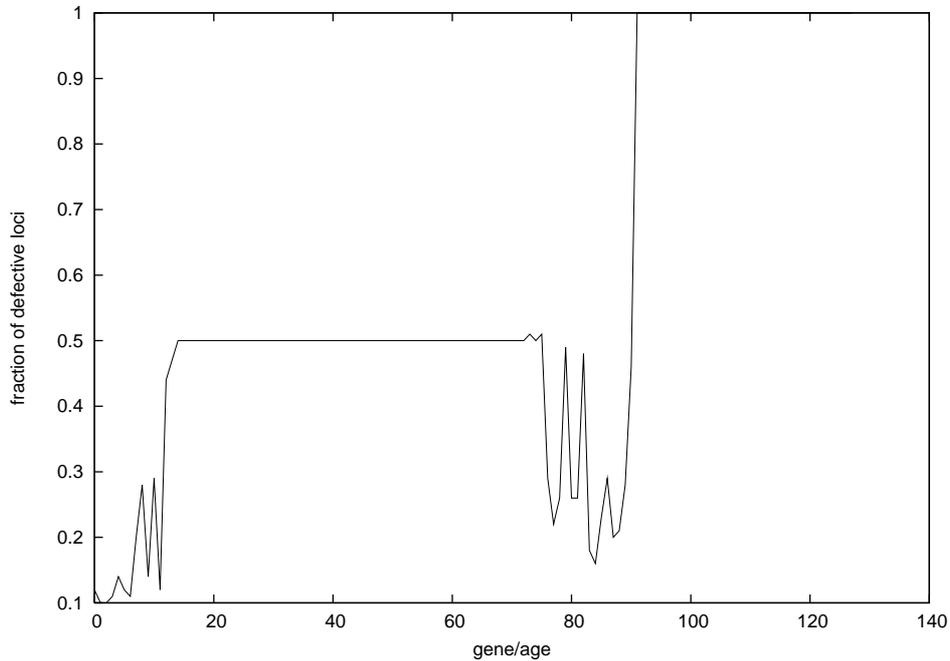}
\caption{The genetic structure of population. L=128, R=80, T=1, C=0.075}\label{f6}
\end{figure}

The critical crossover rate for population without invertion of bits has been found. All parameters of simulations but the crossover rate were constant. Populations switch their strategy of genomes evolution from the complementing haplotypes to the purifying selection at the recombination rate of about 0.075. Under such a regime of crossover rate the distribution of defective genes along the chromosome is characteristic --- the first bits of~the~bitstrings and the bits expressed at the beginning of the reproduction period have lower fraction of defects, it seems that these bits are at least partially under the purifying selection (Fig. \ref{f6}). 

Notice that results shown in Fig. \ref{f5} suggest that all genes expressed before the reproduction age are under the same selection pressure. Nevertheless, the selection pressure weakens for genes expressed during the reproduction period because the defective genes may be transferred to the offspring before the expression of the defective phenotype. If a sequence of expression of bits is like in the standard Penna model (no inversions), then the chromosome is asymmetric --- at one end the bits expressed early during the life span and very strongly watched by selection are located, while at the other end the~``empty space'' can be found. ``Empty'' means that all genes located at this end are defective. There is no genetic information in this space which could help to survive the individuals. Can we find any role for this space?

In the real genomes, if two markers (genes) are physically very close, they are usually inherited together because recombination has to happen just between them to separate them and only in such a case they can be transferred to the gametes separately. It is said that they are linked. If two genes are far away from each other, the probability of recombination between them is higher and they can be inherited independently with higher probability. That is why we have repeated the simulations with inverted part of chromosomes, like in Fig. \ref{f5} close to critical recombination rate. In~the~standard structure of Penna chromosomes (not inverted) the empty space is at the end of chromosome and any recombination in this space has no effect on the probability of disruption of any linked, functional group of~genes. In~the~inverted chromosome, the ``empty space'' is in its central part. Recombination in this part increases the probability of independent transfer of genes located at the ends of chromosome to the gametes and zygotes. 

\begin{figure}
\includegraphics[angle=0,width=\textwidth]{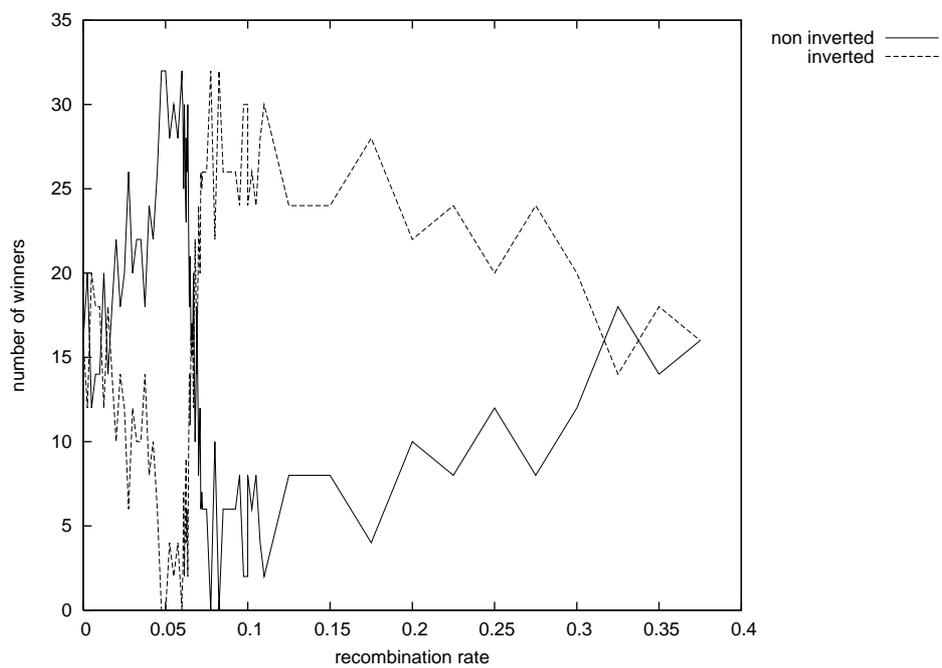}
\caption{Number of winning and losing competitions for populations with standard genome structure and populations with ``inverted'' regions of genomes.}\label{f7}
\end{figure}

To show that the location of this empty space affects the evolutionary value of populations we have simulated the competition of pairs of populations with different structures of chromosomes: without invertion in chromosome and with inverted second part. For the first 400 000 MC steps, populations evolved independently, each in its own environment under independently operating Verhulst factor. After 400 000 MC steps two populations (one with inverted chromosome and the other one without invertion) were put into one environment with doubled $N_{max}$ but with one Verhulst factor operating for both populations, though the populations can not crossbreed, they behave like a different species competing in the same ecological niche. We have performed  such competitions for 32 pairs of populations, each evolving under the same crossover rate. The results are shown in Fig.~\ref{f7}. For low recombination rate - close to 0 as well as for high recombination rate - close to 1, the invertions of chromosomes have no significant influence on the fitness of populations, sometimes the populations with the standard chromosomes are winning and sometimes the populations with the inverted parts of chromosomes are winning. Nevertheless, the structure of chromosomes decides about winning/loosing close to the critical recombination rates. When we increased the recombination rate approaching its critical value, the populations with inverted chromosomes are loosing. Then, the~role is suddenly switched and populations with inverted chromosomes are winning. It can be interpreted that the effective values of recombinations in the inverted chromosomes are higher and they reach their critical value earlier. Since the reproduction potential of population at the critical point is the lowest, the population looses. Under only slightly higher recombination rate the population with inverted chromosomes is already ``above'' the~critical point under purifying selection while the one without invertions just enters the critical range and it is loosing, now. 

\begin{figure}
\includegraphics[angle=0,width=\textwidth]{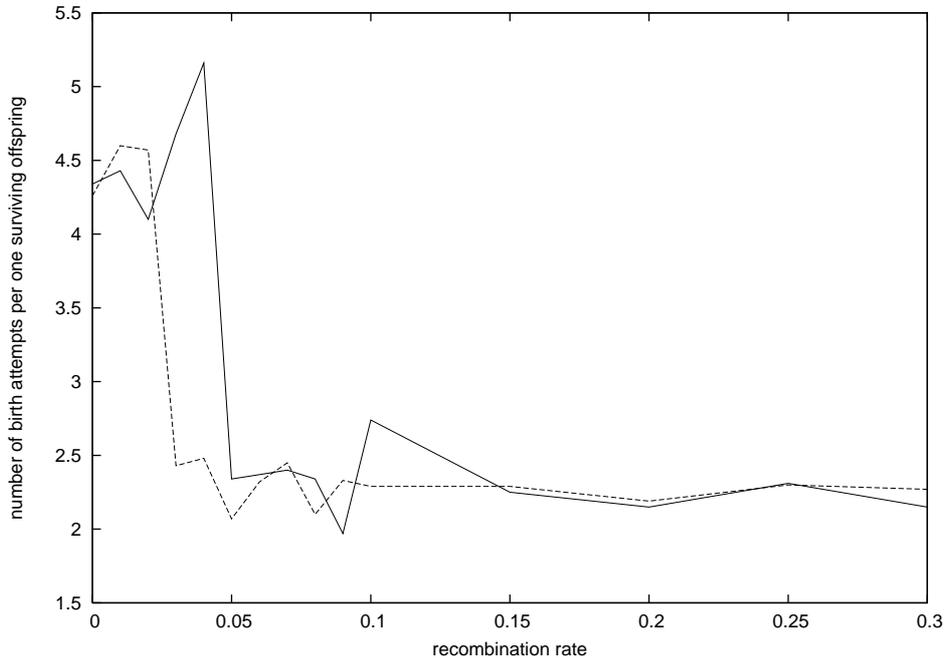}
\caption{Results of~simplified simulations without chronological switching~on bits. Dashed line for the genomes with neutral region in the middle of the bitstrings, solid line for the neutral region at the end of the bitstrings.}\label{f8}
\end{figure}

To support that way of interpretation we have performed the simplified simulations without chronological switching on bits. Individual genemes were composed of two bitstrings 128 bits long. Only 100 of them were under selection, mutations in the rest of them (28) were neutral. These 28 bits were located at the end of chromosome (corresponds to the standard Penna chromosome without invertion) or in the middle of chromosome (corresponds to the chromosome with invertion). The populations evolved for the first 100~000 MCs with sexual reproduction. At each MC step 5\% of individuals were killed randomly and the gap was filled up by offspring generated by the surviving individuals. The newborns survived if their genomes had no loci with both alleles defective. For such populations, we have checked  how the probability of generation the surviving offspring depends on the recombination rate. Results are presented in Fig.~\ref{f8}. The critical value of recombination rate (called phase transition by \cite{Zawierta2}) was lower for populations with neutral region located in the middle of chromosomes. Under the critical recombination rate the probability of generation the surviving offspring is the lowest. Nevertheless, only the slight increase of the recombination rate shifts this population into the purifying strategy with much higher probability of generation the surviving offspring. The population with neutral region located at the end of chromosomes reaches the critical point of recombination (with the lowest probability of the offspring survival) when the population with the neutral part in the middle of chromosome is already under purifying selection. The plots at the Fig.~\ref{f8} show the regions where two populations significantly differ in the reproduction potential and the difference depends exlusively on the recombinational structure of chromosomes. 

Along the natural chromosomes, the recombination events are not evenly distributed . There are some regions with very high recombination frequency (called recombinational hotspots) and some regions with very low frequency of recombination (called recombinational deserts) \cite{Yu}, \cite{Daly}, \cite{Jeffreys}, \cite{Arnheim}. The~``empty space'' in the Penna model chromosome with inversion may be interpreted as recombination hotspot in the middle of chromosome. In the natural chromosomes hotspots do not need to be physically long, they could be just places promoting the recombination. Understanding the distribution of hotspots and its relations to the selection pressure on adjacent genes may be very important for success of genetic manipulations with eukaryotic chromosomes and genomes.

\section{Conclusions}

The results of computer simulations have shown that the distribution of~recombination events along the chromosomes plays an important role in~the~evolution of genomes and populations and it depends on the selection pressure exerted on the genes separated by recombination. Since the Penna model produces the gradient of selection pressure, it seems that it can be used for simulation of self organisation of chromosome structure. 

\section{Acknowledgements}
The work was done in the frame of European programs: COST Action MP0801, FP6 NEST - GIACS and UNESCO Chair of Interdisciplinary Studies,
University of Wroc\l{}aw. Calculations have been carried out in Wr\l{}aw Centre for Networking and Supercomputing (http://www.wcss.wroc.pl), grant\# 102.


\begin{thebibliography}{99}
\bibitem{Ayala} Ayala F.J., J.A. Kiger (1980) Modern genetics, ed. The Benjamin/Cummings Publishing Company, Inc.
\bibitem{Zawierta} Zawierta M., Biecek P., Waga W., Cebrat S. (2007) The role of intragenomic recombination rate in the evolution of population's genetic pool. \emph{Theory Biosci}. 125, 123-132.
\bibitem{Waga} Waga W., Mackiewicz D., Zawierta M., Cebrat S. (2007) Sympatric speciation as intrinsic property of expanding populations. \emph{Theory Biosci}. 126, 53 -- 59.
\bibitem{Fisher} Fisher R.A. (1930) \emph{The genetical theory of natural selection}. Clarendon Press, Oxford
\bibitem{Wright1} Wright S. (1931) Evolution in Mendelian populations. \emph{Genetics}. 16:97-159. 
\bibitem{Wright2} Wright S. (1932) The roles of mutation, inbreeding, crossbreeding and selection in evolution. \emph{Proceedings of the 6th International Congress of Genetics}. 1:356-366
\bibitem{Som} Som C., Bagheri H.C., Reyer H-U. (2007) Mutation accumulation and fitness affects in hybridogenetic populations: a comparison to sexual and asexual systems, \emph{BMC Evolutionary Biology}. 7, 80
\bibitem{Redfield} Redfield R.J. (1994) Male mutation rates and the cost of sex for females. \emph{Nature}. 369, 145-147.
\bibitem{Zawierta2} Zawierta M., Waga W., Mackiewicz D., Biecek P., Cebrat S. (2008) \emph{Phase Transition in Sexual Reproduction and Biological Evolution}. Int. J. Mod. Phys. C. 19 (6) (in press)
\bibitem{Penna} Penna T.J.P. (1995) A bit-string model for biological aging. J. Stat. Phys., 78:1629-1633
\bibitem{scaling} \L{}aszkiewicz A., Cebrat S., Stauffer D. (2005) Scaling effects in the Penna ageing model. Adv. Complex Systems 8:7-14
\bibitem{Moss} Moss De Oliveira S., De Oliveira P.M.C., Stauffer D. (1999) Evolution, Money, War and Computers. Stuttgard-Leipzig: Teubner
\bibitem{Stauffer} Stauffer D., Moss de Oliveira S., de Oliveira P. M. C., S'a Martins J. S. (2006)
\emph{Biology, Sociology, Geology by Computational Physicists}. Amsterdam: Elsevier.
\bibitem{SaMartins} Sa Martins J.S., Cebrat S. (2000) Random deaths in a computational model for age-structured populations. Theory in Bioscien.119:156-165
\bibitem{Aga} \L{}aszkiewicz A., Szymczak Sz., Cebrat S. (2003) Speciation Effect in the Penna Ageing Model. Int. J. Modern Phys. C, 14 (6), 765-774.
\bibitem{Cebrat} Cebrat S., Stauffer D. (2008) Gamete recognition and complementary haplotypes in sexual Penna ageing model, Int. J. Mod. Phys. C. Vol. 19 (2) 259 -- 265,
\bibitem{Yu} Yu A., et al. (2001) Comparison of human genetic and sequence-based physical maps. Nature 409:951-953.
\bibitem{Daly} Daly M.J., Rioux J.D., Schaffner S.F., Hudson T.J., Lander E.S. (2001) High-resolution haplotype structure in the human genome. Nature Genet. 29:229-232.
\bibitem{Jeffreys} Jeffreys A.J., Kauppi J., Neumann R. (2001) Intensely punctate meiotic recombination in the class II region of the major histocompatibility complex. Nature Genet. 29:217-222.
\bibitem{Arnheim} Arnheim N., Calabrese P., Nordborg M. (2003) Hot and Cold Spots of Recombination in the Human Genome: the Reason We Should Find Them and How This Can Be Achieved. Am. J. Hum. Genet. 7:5-16.

\end{thebibliography}
\end{document}